# Self-powered, flexible and room temperature operated solution processed hybrid metal halide p-type sensing element for efficient hydrogen detection

E. Gagaoudakis[1,2, †], A. Panagiotopoulos[3,5 †], T. Maksudov[3,5 †], M. Moschogiannaki[2,3], D. Katerinopoulou[1,2], G. Kakavelakis[3,4], G. Kiriakidis[1,2], V. Binas[1,2], E. Kymakis[5] and K. Petridis[6, *]

[1]*Department of Physics, University of Crete, Heraklion, Greece*
[2]*Institute of Electronic Structure and Laser, Foundation for Research and Technology Hellas, 100 N. Plastira str., Vassilika Vouton, 70013 Heraklion, Crete, Greece*
[3]*Department of Materials Science and Technology, University of Crete, GR-71003, Heraklion, Greece*
[4]*Cambridge Graphene Centre, University of Cambridge, 9 JJ Thomson Avenue, Cambridge CB3 0FA, UK*
[5]*Department of Electrical & Computer Engineering, Hellenic Mediterranean University, Estavromenos P.B 1939, Heraklion, GR-71 004, Crete, Greece*
[6]*Department of Electronic Engineering, Hellenic Mediterranean University, Romanou 3, Chalepa, 73100, Chania, Crete, Greece*
[†]these authors contributed equally to the preparation of the manuscript

**KEYWORDS.** lead halide perovskites, self-powered hydrogen sensing element, Solution Processed, Room Temperature Sensing, p-type sensing elements, sensitivity, flexible substrate

## Abstract

Hydrogen ($H_2$) is a well-known reduction gas and for safety reasons is very important to be detected. The most common systems employed along its detection are metal oxide-based elements. However, the latter demand complex and expensive manufacturing techniques, while they also need high temperatures or UV light to operate effectively. In this work, we first report a solution processed hybrid mixed halide spin coated perovskite films ($CH_3NH_3PbI_{3-x}Cl_x$) that have been successfully applied as portable, flexible, self-powered, fast and sensitive hydrogen sensing elements, operating at room temperature. The minimum concentrations of $H_2$ gas that could be detected was down to 10 ppm. This work provides a new pathway on gases interaction with perovskite materials, launches new questions that must be addressed regarding the sensing mechanisms involved due to the utilization of halide perovskite sensing elements while also demonstrates the potential that these materials have on beyond solar cell applications.

## Introduction

Hydrogen (H$_2$) gas is expected to be a green (no emissions) and renewable energy source (with high heat combustion 142 kJ/g, minimum ignition energy 0.017 mJ and high flammable range up to 75%) for many applications such as glassmaking, semiconductor processing, biomedical applications, seismic surveillance, fuel cells, automobiles, power generators and aerospace (liquid H$_2$ already been used for rocket fuels) [1,2]. In the near future it could be used as a city gas or as a fuel to power cars in the same way as natural gas is leveraged. However, H$_2$ is an extremely dangerous gas since it is odourless, colourless, and highly flammable, with high burning velocities and its leakage poses explosion hazards (a lower explosion limit (LEL) at 40000 ppm) [3]. So, it is essential (see Figure 1) to be detected reliably and fast, in low concentrations preferably below 100 ppm, in order to monitor possible leakage during storage and transport, for safety protection reasons [4-7].

A H$_2$ sensor could be a transducer that converts a variation of physical or chemical characteristic into an electrical current. A number of applications, including gas chromatography, mass spectrometry, thermal conductivity, laser-induced breakdown spectroscopy, scanning photoelectrochemical microscopy, and gas sensors, have been employed to detect hydrogen gas [8-14]. Among them, the hydrogen gas semiconductor-based sensors are being studied for their small size; low power consumption; high accuracy, reliability; fast response; and reliable / low cost fabrication processes [15]. Subject to the signal monitored, the H$_2$ sensing elements can be divided into (a) resistance based; (b) work function based; (c) optical; and (d) acoustic elements. Figures of merit of an ideal sensing element are (a) the sensitivity to various targeted gas agents; (b) the selectivity between various targeted gases; (c) the fast response when exposed in the environment of the targeted gas; (d) reversibility to its initial pristine stage before the exposure to the targeted agent; (e) the efficient detection of the signal generated as a result of the interaction of the sensing element with the targeted gas; (f) the low cost and facile fabrication process; (g) stability and long life time (h) the operation at low temperature (ideally at room temperature), without the need of an external trigger (temperature or UV light) to provide sensing abilities.

In this paper, we will be focused on the resistance-based sensing element technique and demonstrate a very promising performance towards H$_2$ gas sensing. The motivation originated from the fact that the most common employed H$_2$ sensing materials (metal oxides

– $SnO_2$, $ZnO$, $TiO_2$, $Nb_2O_5$, $In_2O_3$, $FeO$, $NiO$, $Fe_2O_3$, $Ga_2O_3$, $MoO_3$, $V_2O_5$, $WO_3$) do not contain all the figures of merit of an ideal sensing element, despite their appealing characteristics: small size, high sensitivity, high repeatability and simplicity to use. Often, they suffer from some serious drawbacks such as (a) high operation temperatures (operation at room temperature has shown weak response to low concentration $H_2$ or have long response-recovery time. Moreover operating at high temperatures is very risky especially when the targeted gas is an explosive gas such as $H_2$); (b) they need a pre-treatment with UV light in order to become conductive; and (c) their fabrication is complicated and expensive (e.g. rf sputtering, dc sputtering, high vacuum facilities, pulsed laser deposition [16-21].

Hybrid metal halide perovskite materials are the new star materials in the solar cell technological field and not only. They are of the type $ABX_3$ where, A is an organic cation usually methylammonium ($MA^+$), formamidinium ($FA^+$) etc. or an inorganic cation such as $Cs^+$, $Rb^+$ and $K^+$, B is a metal such as $Pb^{2+}$, $Sn^{2+}$, while the anion X is a halogen such as $I^-$, $Br^-$, $Cl^-$ or a mixture thereof. The very first report on semiconductor halide perovskite was via observing photoconductivity in the all inorganic $CsPbX_3$ systems at late 1950'by Moller [22]. Later on, perovskite was used as absorbing material in photovoltaic applications by Kojima et al. in 2009 [23]. In the last decade, the efficiency of perovskite based solar cells have been skyrocketed from 3.8% to higher than 25% [24]. The hybrid lead halide perovskites, such as the mixed halide one studied here ($CH_3NH_3PbI_{3-x}Cl_x$), have recently attracted the scientific communities' attention as one of the most promising solar light energy harvesting materials due to their direct bang gap, long diffusion charge carrier lengths, large absorption coefficients, long carrier lifetime and large carrier mobility [25-27]. The impressive impact of perovskite in solar cell technology, have attracted an intensive research interests towards the applications of perovskites beyond solar cells such as in lasers [28] in light emitting diodes [29,30] and photo-detectors [31,32]. However, one of the key issues to be solved in order to boost their commercialization is related to their stability. Halide perovskites are very sensitive to polar gases and vapours, as, exposure to such elements deteriorates substantially and very fast the perovskite devices' performance. The conversion of this disadvantage to an advantage reflects the introduction of perovskite-based gas sensing elements. The sensitivity of perovskite to environmental gases is an opportunity to convert a drawback to an opportunity [33]. Recently, halide perovskites have been explored and demonstrate their competence over the existed technologies, as potential sensing elements from gas molecules to X-ray photons [34-38]. Increasingly new publications regarding inorganic perovskites applications in $H_2$ sensing, have started to appear [39,40] (Figure 1) showing the potential of this family of materials towards hydrogen sensing. However, all reported inorganic perovskite materials need to operate at very high temperatures (of few hundred Celsius) in order to function as hydrogen sensors. In this paper, we introduce for the first time a solution processed hybrid metal halide perovskite film ($CH_3NH_3PbI_{3-x}Cl_x$), that requires much simpler fabrication and deposition techniques & facilities than the previously reported for other $H_2$ sensing element materials. Moreover, the performance characteristics of the demonstrated sensing element are not inferior to many of figures of merit of the other technologies reported. The perovskite based sensing element presented here, demonstrated a very promising performance characteristics towards hydrogen sensing, i.e.: (a) operation at room temperature; (b) no requirement of an external optical signal to be switched on prior to its exposure into a hydrogen environment; (c) fast response time (few secs); (d) detection sensitivity with resolution down to 10 ppm; and (e) compatibility with flexible substrates. The aforementioned features make hybrid mixed halide perovskite semiconductors competitive (and beyond the simpler and of lower cost fabrication and deposition techniques) candidates to other already demonstrated in hydrogen sensing materials such as metal oxides and metals [4]. More particular compared to metal oxides (e.g. $SnO_2$, $ZnO$, $TiO_2$, $FeO$, $Fe_2O_3$, $NiO$, $Ga_2O_3$, $In_2O_3$, $Sb_2O_5$, $MoO_3$ and $WO_3$) that request high temperatures to operate as hydrogen sensing elements, the demonstrated hybrid perovskite film operate in room temperature reducing a lot the consuming power. To highlight that the performance of both systems regarding the reaction times (of the order of few seconds) and minimum detection limits are similar (of the order of 10 ppm). Regarding their advantages compared to metallic resistors (e.g. palladium & platinum metals) (a) request simpler deposition techniques (e.g. spin coating) compared to more complicated and expensive techniques the metallic resistors request: vacuum evaporation, electrode position, sputtering and pulsed laser deposition; (b) do not suffer from mechanical degradation when exposed to hydrogen environment and thus do not request complicated metallic alloys to address the issue of the mechanical degradation. To add that hybrid perovskite semiconductors, show similar minimum detection limits as metallic resistors.

The studied sensing films demonstrated a p-type semiconductor behaviour (attributed to its stoichiometry) and thus under $H_2$ gas (reducing gas) exposure their electrical resistance was enhanced upon $H_2$ exposure. However, its pristine electrical properties were restored very fast (within few secs) after the removal of the $H_2$ gas. This work is believed to set the framework for research of low temperature operated, efficient and of low cost new conductometric (resistance measurement), compatible with flexible electronic industry halide perovskite $H_2$ sensing elements and systems. This is essential for various applications ranging from the energy sector to aerospace industry.

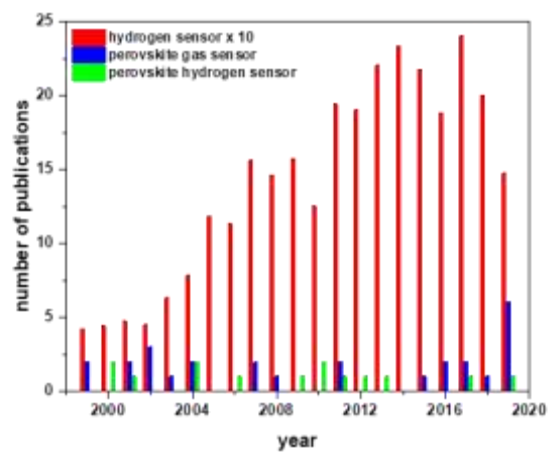

**Figure 1:** Number of publications regarding $H_2$ sensing and applications of perovskites as gas sensors and $H_2$ sensors (Source Scopus)

**Methodology**

To fabricate the $CH_3NH_3PbI_{3-x}Cl_x$ precursor solution, methylammonium iodide (MAI, purchased by Dyesol) with lead (II) chloride ($PbCl_2$, 99.999% purchased by Sigma Aldrich) were mixed in a molar ratio of 3:1 within a anhydrous N,Ndimethylformamide (DMF) solvent. The resulted solution (with 40 wt% concentration) was continuously stirred on a hot plate for 12 hrs at 70 °C. Afterwards, the precursor solution was cooled down at room temperature and deposited (approximately 20 μl filtered precursor solution) onto the sensing element electrodes template (purchased from DropSens). The latter and prior to the spreading of the solution precursor, was UV ozone cleaned to remove any hazards (e.g. organic contaminants) that can lower its hydrophilicity. The electrode substrate was glass made and on top of the glass two interdigitated platinum electrodes were pre-patterned. The distance between the electrodes was 5 μm. The deposited solution was spin-coated at 4000 rpm for 45 seconds.

Afterward, the spin-coated precursor solution over the electrodes was thermally annealed at 100 °C for 75 minutes to form the $CH_3NH_3PbI_{3-x}Cl_x$ hybrid perovskite semiconductor. The entire processing was conducted inside a nitrogen-filled glove box with $O_2$ and $H_2O$ concentrations below 0.1 ppm. Prior to the testing of the sensing properties of the glass/Pt/$CH_3NH_3PbI_{3-x}Cl_x$ sandwich elements, the latter were fully characterized as a function of their electronic, morphological, structural properties. Prior to the sensing testing the glass/Pt/$CH_3NH_3PbI_{3-x}Cl_x$ system was exposed under simulated solar light (A.M 1.5G at 100mW/cm$^2$) in ambient conditions (~45% relative humidity). The film roughness was examined using Atomic Force Microscopy (AFM), by employing a Park XE-7 instrument in tapping mode. The total scan area was set to 50 μm x 50 μm and the scan rate was fixed at 0.3 Hz. The successful crystallization of the hybrid $CH_3NH_3PbI_{3-x}Cl_x$ perovskite semiconductor was assessed using a D/MAX-2000 X-Ray diffractometer under monochromatic Cu Kα irradiation (λ=1.5418 Å) at a scan rate of 4° min$^{-1}$. Whereas the grain size of the crystallized perovskite film was evaluated through Scanning Electron Microscope (SEM, using the JEOL JSM-7000F) measurements.

All the hydrogen sensing measurements occurred within a homemade gas test chamber under dark conditions in order to lower the rates of photoexcitation that contributes to dark current. Prior to the sensing test measurements that occurred, the electrical conductivity of our elements was tested. A potential difference of one Volt was applied across the two platinum electrodes of glass/Pt/$CH_3NH_3PbI_{3-x}Cl_x$ systems and the induced generated current was measured using a Keithley 6517A multi-meter set-up. All these initial measurements were taken at room temperature and under a pressure of 1.6 mbar.

After the baseline of the conductivity of our sensing elements was measured, the $CH_3NH_3PbI_{3-x}Cl_x$ sensors were exposed for five minutes to hydrogen gas at a constant flow of 500 sccm (standard cubic centimetres per minute), while the pressure in the chamber was kept constant at 120 mbar, leading to a decrease of current. The hydrogen concentrations that the detector was exposed were 100, 75, 50, 25, and 10 ppm. The sensing performance of our sensing elements could not be assessed for hydrogen concentrations higher than 100 ppm for safety reasons. The $CH_3NH_3PbI_{3-x}Cl_x$ sensor was exposed for time intervals of five minutes at each concentration, while another five minutes was given to the sensor to relax to its steady-state conditions.

**Results**

**1. Structural, Morphological and Optical Analysis**

A 300 nm thick $CH_3NH_3PbI_{3-x}Cl_x$ film was fabricated onto a glass substrate with prepatterned electrodes; the interdigitated electrodes made of platinum had distance between them of 5 µm. It must be highlighted at this point that the majority carriers within the perovskite semiconducting film are subject to the stoichiometry of the $PbI^2/MA^+I$ precursor ratio; in our case this ratio was less than one and the film demonstrated a p-type semiconducting behaviour [41]. The entire fabrication process was performed inside a glovebox under nitrogen atmosphere.

The film's surface morphology controls the number of the provided interaction pathways that the targeted gas molecules can interact on with the sensing element; increased roughness, facilitates the gas molecules adsorption by providing longer diffusion lengths within the active material. The surface film's morphology was revealed using Atomic Force Microscopy (AFM) measurements (Figure 2). AFM patents were taken before (Figure 2a) and after the exposure to $H_2$ (Figure 2b). A striking change in the AFM images was observed, with the roughness to reduce almost to its half value after the exposure to 100 ppm of hydrogen gas. The reason of this morphological change is thought to relate to the interaction of H-molecules with the perovskite surface species; however, the exact cause needs further investigation before it can be attributed to a specific factor. Subsequently, a slight decrease in the measured current (of the order of few nA during the measurement cycles) was attributed to the resulted smoother surfaces; the latter provide shorter percolations paths to allow the $H_2$ molecules to interact with the perovskite platform.

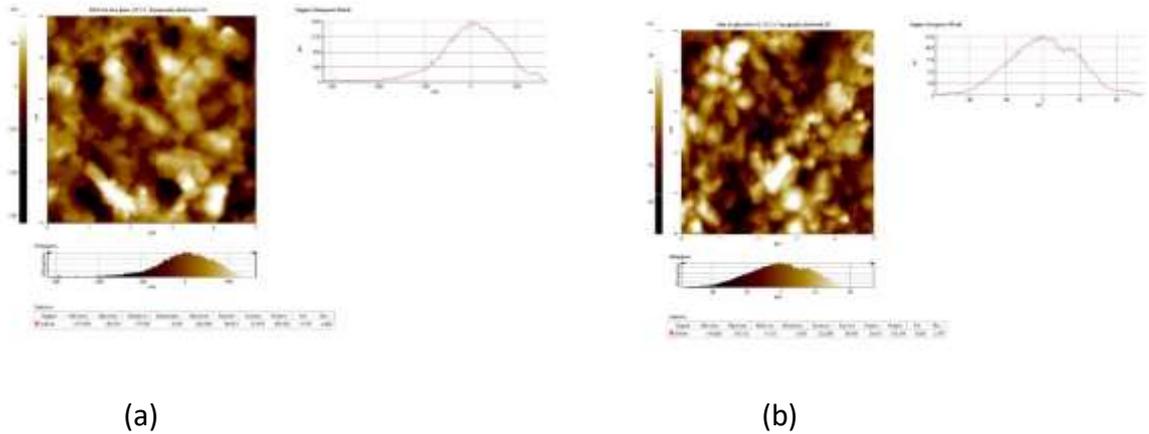

|        (a)         |        (b)         |

**Figure 2:** AFM image of the surface morphology of the MHP film; the RMS value was 69.27 nm before the $H_2$ exposure (a) whereas after prolonged exposure the morphology has changed to 36.70 nm (b)

The Scanning Electron Microscopy (SEM) images of the employed mixed halide films reveal the well-shaped formed grains on its surface and the film's porosity. The existence of a porous surface on the film, facilitates the penetration and the evacuation paths of the hydrogen atoms through it (Figure 3). The good conductivity of the film is very much linked with the grain size; using the Scherrer formula their size was measured of the order of 250 nm. The SEM images before and after the exposure to the $H_2$ gas revealed no substantial changes on the grain size.

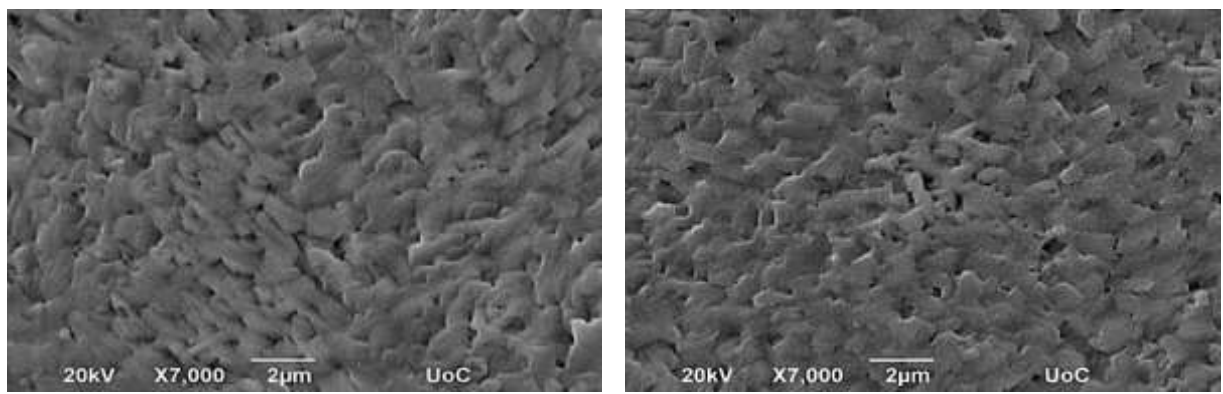

**Figure 3:** SEM image of the surface of the MHP employed as sensing element for hydrogen before (left image) and after (right image) exposure to hydrogen. No changes in the morphology were observed. Using the Scherrer equation the grain size was calculated to be of the order of 250 nm

The film's crystal structure was studied using the X-ray diffraction (XRD) technique. The successful crystallization of the spin coated perovskite film is depicted in the acquired XRD image (Figure 4). The crystal directions of the first (110) and the second (220) crystallographic plane can be seen, at ~14.2° and ~28.5° respectively, confirming the, as expected, cubic perovskite phase [42]. No compositional changes were observed before and after the $H_2$ to hydrogen gas molecules (at 100 ppm for more than one hour), showing the sensing potential and stability of the fabricated films. This is a clear indication that the interacted $H_2$ molecules just adsorbed and de-adsorbed through the perovskite-based template.

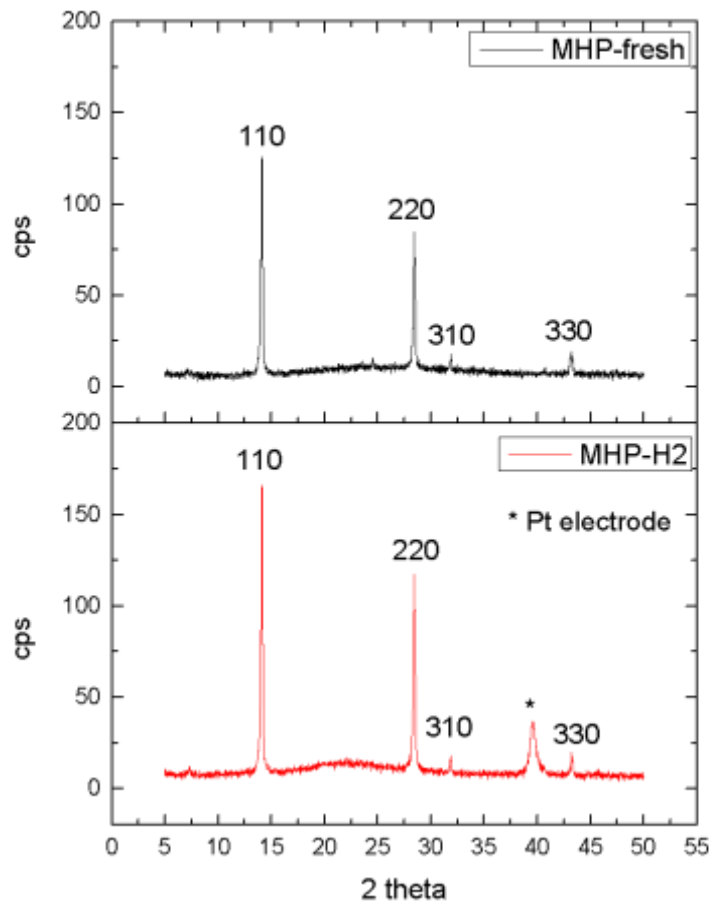

**Figure 4:** XRD images of the mixed halide perovskite films before and after its exposure to 100 ppm hydrogen gas. The results showed no structural changes after the interaction of the sensing element with the hydrogen. The XRD of the non-exposed halide perovskite film was taken on a glass substrate without electrodes

## 2. Sensing Properties of the $CH_3NH_3PbI_{3-x}Cl_x$ thin films

The sensing functionality of the films on $H_2$ was examined by electrical measurements at room temperature, carried out in a home-made set-up. We exposed the aforementioned perovskite films to various $H_2$ gas concentrations up to the maximum of 100 ppm for safety reasons. Interruption the $H_2$ gas flow, restored the films' conductivity to their initial levels. The excellent electrical properties (long diffusion lengths, high charge mobility and excellent mobility to charge lifetime product), the interdigitated electrodes (Pt electrodes with distance of the order of 10 μm) setup and the application of a forward bias across the electrodes (0.5 and 1 V) allowed us to register (a) the current across the film (of the order of μA); and (b) its modulations when the film was exposed to $H_2$ environment. The report of such high currents (in μA range) through the film is an indication of its excellent electrical properties as a result of the long grain's size and high carrier's diffusion length of the $CH_3NH_3PbI_{3-x}Cl_x$ perovskite layer employed as sensing material. The demonstrated sensing elements were self-powered thus there was no the need of any external assistance such as UV irradiation or heating in high temperatures. $H_2$ molecules operated as a reducing gas, and since the exhibited perovskite film is a p-type semiconductor, lowering of the current was observed, as expected, caused by the adsorption of the $H_2$ molecules (the opposite behaviour than n-type semiconductors). Figure 5 depicts the current modulation for (a) various exposure times (one and five minutes); and (b) under different forward biases 0.1, 0.5 and 1 Volt. The O.1 V acquired measurements were ignored since provided very low currents and low sensitivities.

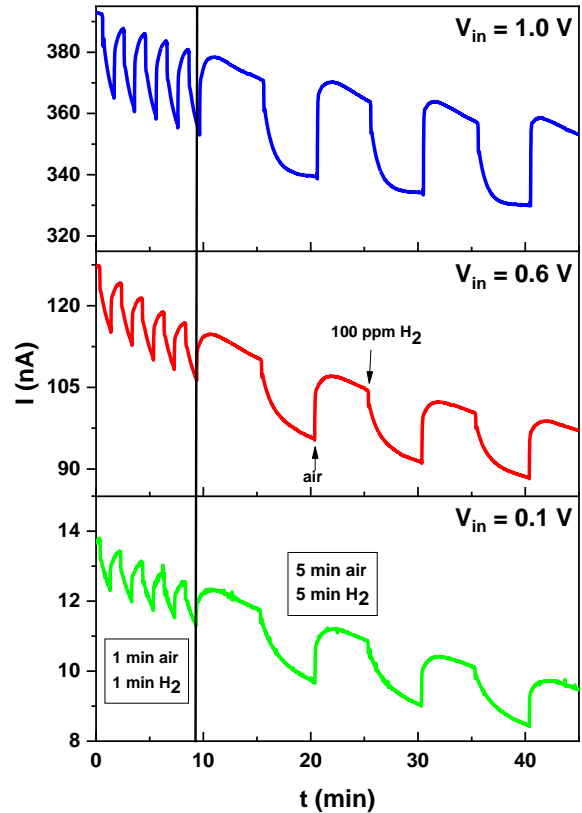

**Figure 5:** Dependence of the sensitivity (with the external bias) and demonstration of the reproducibility & stability of the solution processed $CH_3NH_3PbI_{3-x}Cl_x$ sensing elements as a function of the operational time (under exposure at multiple cycles of synthetic air-100 ppm $H_2$) and the applied external bias (1V, 0.5V and 0.1 V)

During the sensing measurement process, the modulation of the current flowing through the film was recorded, as the $H_2$ gas in the synthetic air was inserted and stopped, following several cycles. The film demonstrated sensing reversibility as this is illustrated in Figure 6. Interruption $H_2$ gas flow lead to the recovery of the prior sensing element electrical resistance. The reproducibility of the acquired results, supports the credibility of our sensing element performance and reliability. The sensing film demonstrated a reversible sensing behaviour under various hydrogen gas concentrations: 100 ppm, 75 ppm, 50 ppm, 25 ppm and 10 ppm. As a result, the reducing gas causes an increase in the film resistance, reconfirming its p-type character.

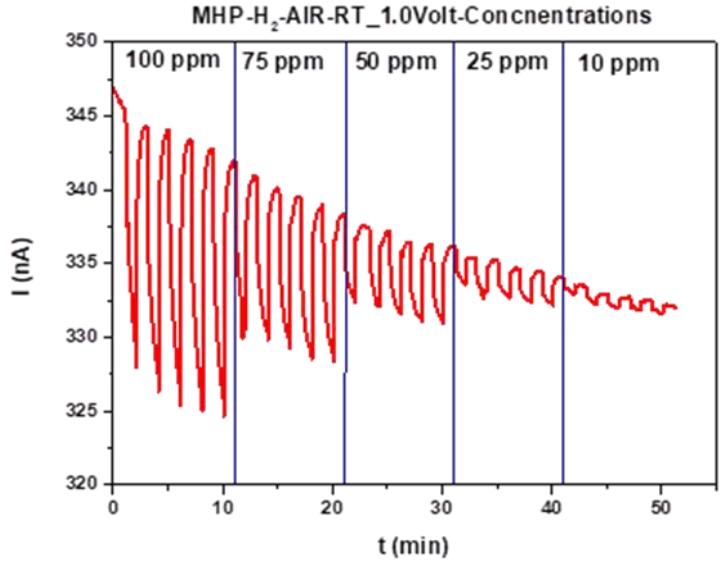

**Figure 6:** Reversible current modulation through 300 nm thick mixed halide perovskite films under exposure (for five minutes) in various hydrogen concentrations. As was expected the sensitivity drops with the concentration of the hydrogen atoms within the chamber. The minimum detection limit was 10 ppm of $H_2$

As expected, our perovskite sensing platform demonstrated different sensitivity at different targeted gas concentrations (Figure 6). This, in combination with the abrupt increase of the resistance, was attributed to the $H_2$ molecules adsorption into the perovskite surface. Furthermore, as the need for low temperature gas sensors is becoming a necessity and moreover for safety reasons in the case of the highly flammable $H_2$ gas, all the measurements were taken at room temperature although sacrificing part of the response signal obtained at elevated temperatures. All the measurements were recorded until the current through the sample had reached its lowest value. This process duration was lasted for approximately five minutes after which no further substantial changes were observed. The sensing measurement was repeated under each hydrogen gas concentration for five times (five cycles) –Figure 6, showing excellent reproducibility. The sensing ability was assessed with the measurement of (a) the sensitivity (S); and (b) the response & recovery times of the film. The sensitivity parameter is calculated using the following formula (1) [43]

$$S(\%) = \frac{|I_{air} - I_{gas}|}{I_{air}} 100\% \quad (1)$$

where $I_{air}$ denotes the electrical current of the sensor before the exposure into the $H_2$ environment, $I_{gas}$ the resistance of the sensor after ten minutes to hydrogen. The response ($t_{resp.}$) and the recovery ($t_{recov.}$) times were calculated as the times take the measured current to reach the 10% of its maximum value under $H_2$ exposure and 90% of its maximum value under synthetic air environment ($H_2$ gas flow has been interrupted), respectively. The response and recovery times demonstrated almost identical average values, 45 and 35 seconds, respectively. The dependence of sensitivity with the $H_2$ concentration is depicted on Figure 7.

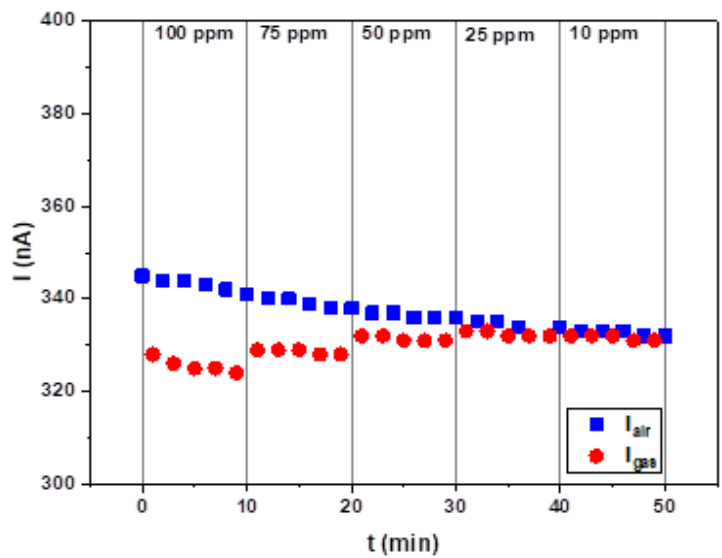

**Figure 7:** Sensitivity dependence on the concentration of the $H_2$ atoms

Table 1 presents all the calculated values, extracted from the measurements depicted on Figure 7, under the different $H_2$ gas concentrations taken. All these values have been plotted and illustrated on Figure 8.

| $H_2$ concentration (ppm) | $I_{air}$ (nA) | $I_{gas}$ (nA) | $\Delta I$ (nA) | S (%) | $t_{resp}$ (s) | mean $t_{resp}$ (s) | $t_{rec}$ (s) | mean $t_{rec}$ (s) |
|---|---|---|---|---|---|---|---|---|
| 100 | 343.6 | 325.6 | 18 | 5.2 | 58.50 | | 20.40 | |
| 75 | 339.6 | 328.6 | 11 | 3.2 | 44.58 | | 23.04 | |
| 50 | 336.8 | 331.4 | 5.4 | 1.6 | 34.56 | 45.14 | 42.48 | 35.40 |
| 25 | 334.8 | 332.4 | 2.4 | 0.7 | 49.56 | | 30.00 | |
| 10 | 332.6 | 331.6 | 1 | 0.3 | 38.52 | | 61.08 | |

**Table 1:** Calculated values of the sensitivity (S), response and recovery times of the demonstrated perovskite films under various $H_2$ gas concentration exposures

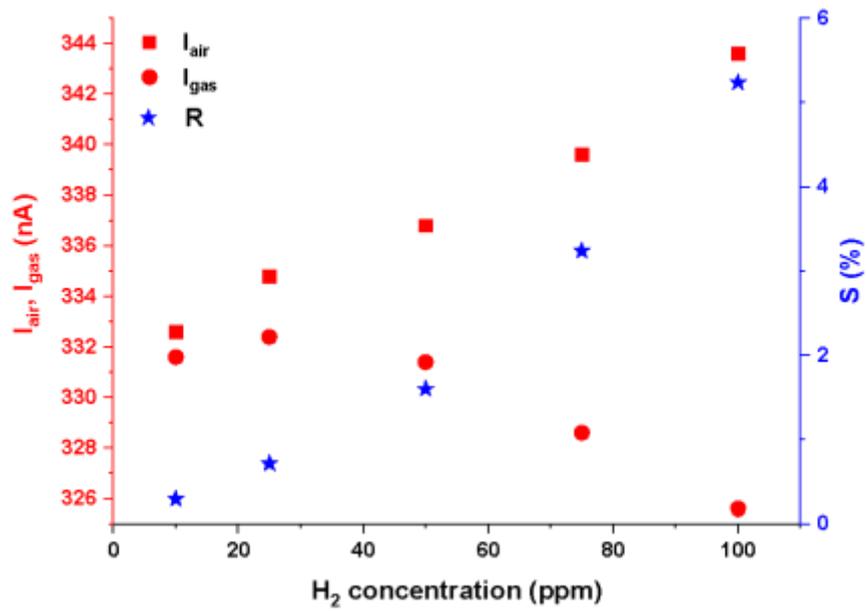

**Figure 8**: Mean values of electrical current after exposure to $H_2$ ($I_{gas}$) and synthetic air ($I_{air}$) and response of sensor as function of $H_2$ concentration. Sensitivity increased with the concentration of the $H_2$ molecules

The fast sensing response of the exhibited films allowed the distinction of the various $H_2$ concentrations. You can see in Figure 10 below, that the maximum exposure time was 10s.

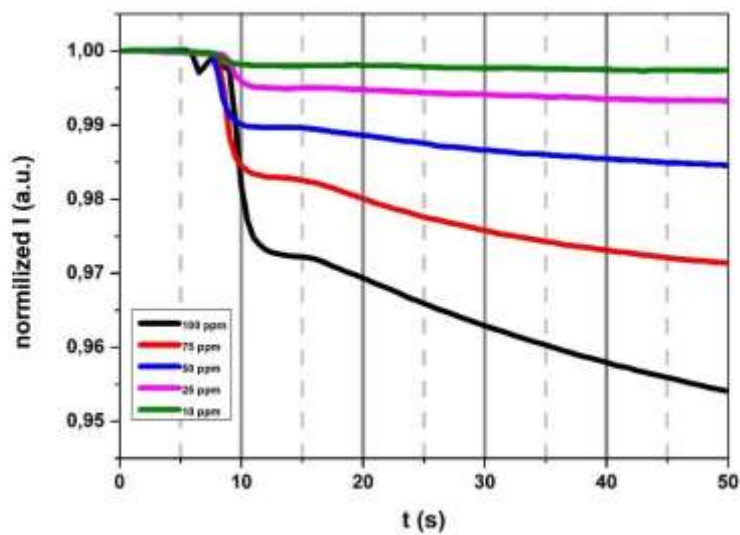

**Figure 9:** Normalized current with time of exposure at different concentrations of $H_2$

The absence of any phase transformation and impurity formation, upon exposure to a $H_2$ atmosphere, shows the non-chemical interaction of the hydrogen with the halide perovskite platform. This is also supported from the reversibility of the electrical properties of the perovskite-based film exhibited after the removal of the $H_2$ gas molecules from the chamber. It is crystal clear that the $H_2$ molecules are adsorbed through the porous of the perovskite film and bond loosely with its crystal template close to the surface; this can be interpreted by the small response and recovery times the film showed. The increase of the films' resistance under the $H_2$ exposure (a reduction gas) indicate the introduction of electrons to the halide perovskite platform. These electrons recombine with the holes (majority charge carriers in the p-type semiconductor) and result in the lowering of the current through the film. We attribute the donation of the electrons to the perovskite platform to the following mechanism: Firstly, neutral oxygen molecules (the most probable from the synthetic air existed within the vacuum chamber) are adsorbed into the perovskite platform and convert into oxygen ions: $O_2^-$, $O^-$ and $O^{2-}$ ions by attracting electrons from the semiconductor valence band. Subsequently, the hydrogen atoms interact with these oxygen ions, producing water with the simultaneous release of electrons. The latter recombine with the perovskite holes in the valence band and result in the observed enhancement of the film's resistance. When hydrogen is desorbed, the surface of the material continues to adsorb oxygen from the environment to generate oxygen ions. At this time, the resistance of the material returns to the base value. However, the XRD measurements showed no compositional changes (the formation of water molecules should lead to a severe degradation and decomposition of the halide perovskite). This does not occur even though the films were exposed to $H_2$ environment for more than an hour. This is an open question. We conclude that further studies and theoretical modelling should be done to support further the aforementioned operational mechanism for the $H_2$ sensing[25].

It is noted that after a number of successive cycles, the current did not fully return to its initial value. This is attributed possibly to residual adsorbed $H_2$ molecules due to slight changes in the film's morphology as the AFM measurements revealed. The observation of no compositional change after the exposure to $H_2$ molecules is attributed to the good crystallinity of the perovskite film that reflects film's good stability. Moreover, the sensing properties in successive hydrogen/synthetic air switching cycles of the mixed halide films was tested after three weeks of storage under vacuum within the measurement chamber demonstrating remarkable sensing element durability. The results were encouraging, with the sensing element to demonstrate a good stability (see Figure 10), as the sensitivity and the detected currents remained almost the same after three weeks of storage.

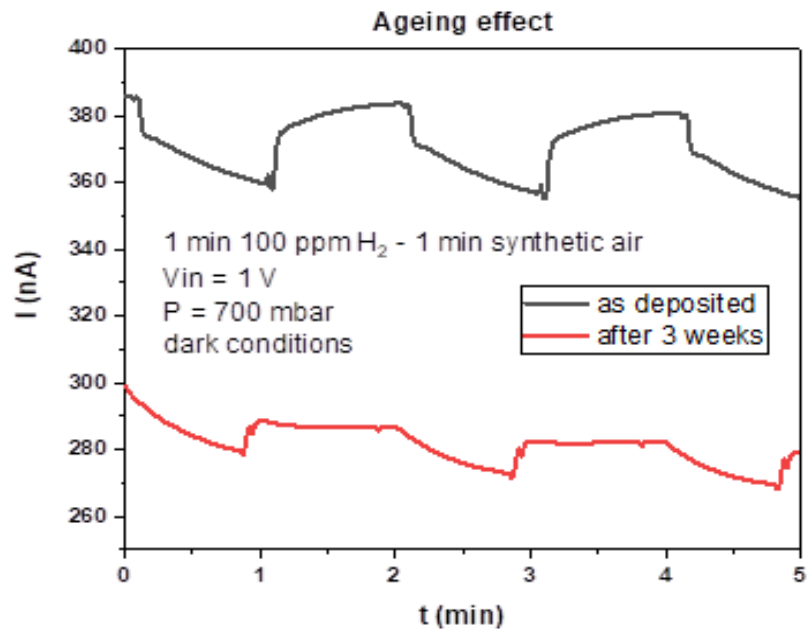

**Figure 10:** Electrical current variation of MHP sensor after exposure at 100 ppm $H_2$ and synthetic air at input voltage of 1.0 V. The measurements were taken for as deposited samples as well as after three weeks

The ability to distinguish different gases was also demonstrated by the fabricated films. The same film exposed to the $H_2$ gas, was exposed to ozone molecules (Figure 11a and b). The latter lowered the resistance of the films. The opposite result in the electronic properties of the same sensing platform the two different gases produced, accompanied with the different calculated sensitivities (in the case of ozone exposure the perovskite film demonstrated much better sensitivities), revealed the ability of the mixed halide perovskite films to distinguish two different gases.

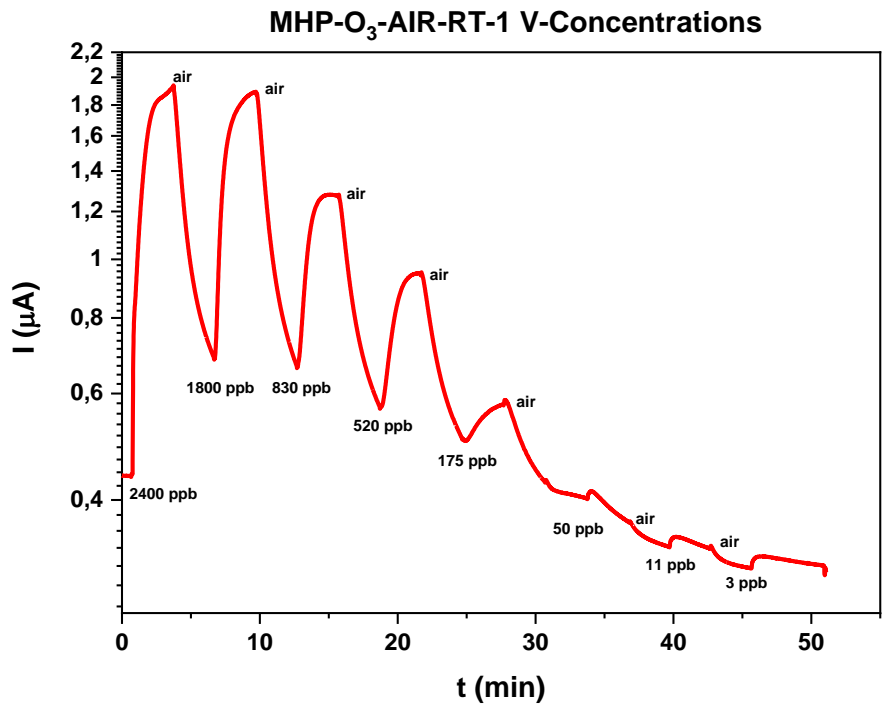

(a)

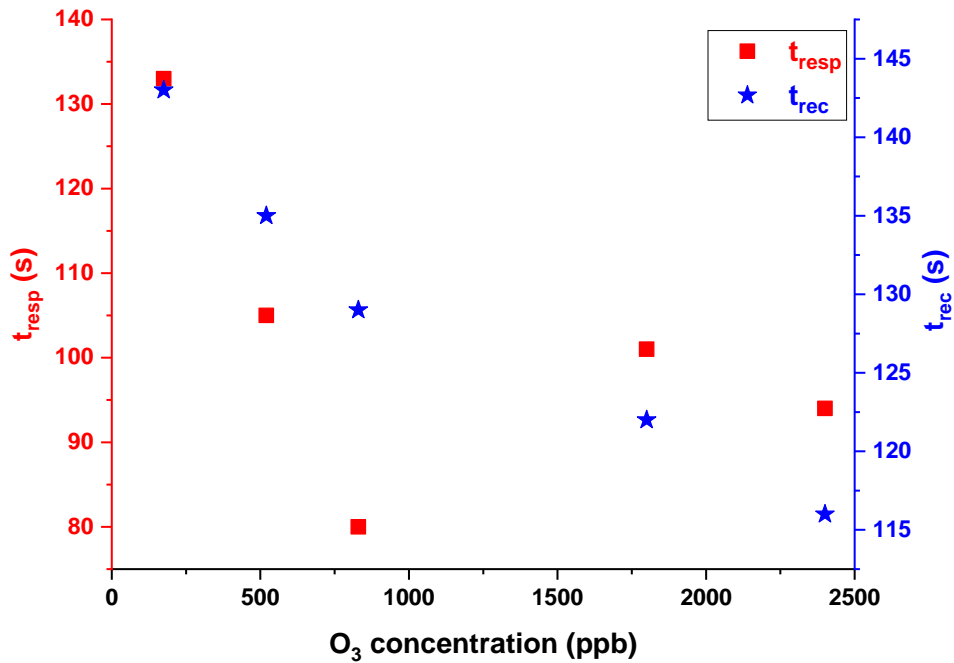

(b)

**Figure 11:** The same film measured ozone molecules; the latter operated as oxidizing gas with the perovskite platform showing the selectivity of our sensing element compared to the reductive behaviour to the same platform of the $H_2$ molecules. **(a)** The modulation of electrical current under exposure to various ozone concentrations. **(b)** The impact of the various ozone concentrations to the sensing times of the perovskite film

The solution processability and the crystallization of the employed perovskite in low temperatures, permitted the deposition of sensing active onto flexible e.g. PET substrates where similar electrode patterns with these films on rigid substrates have been printed. The film was exposed to $H_2$ environment under (a) no bending; and (b) bending conditions. In Figure 12a-c is depicted the sensing performance under bending conditions (Fig. 12b); and the two extreme cases of no bending and bending close to 180° (Fig. 12c); in both cases the modulation of current due to $H_2$ interaction with the perovskite template could be detected and be measured. This is the first time that a report on the flexibility of the tested material as sensor is made and these results are very promising. The sensitivity recorded was comparable (around 7% under bending and exposed to 100 ppm of hydrogen molecules) with the one of the mixed halides films deposited onto glass substrates.

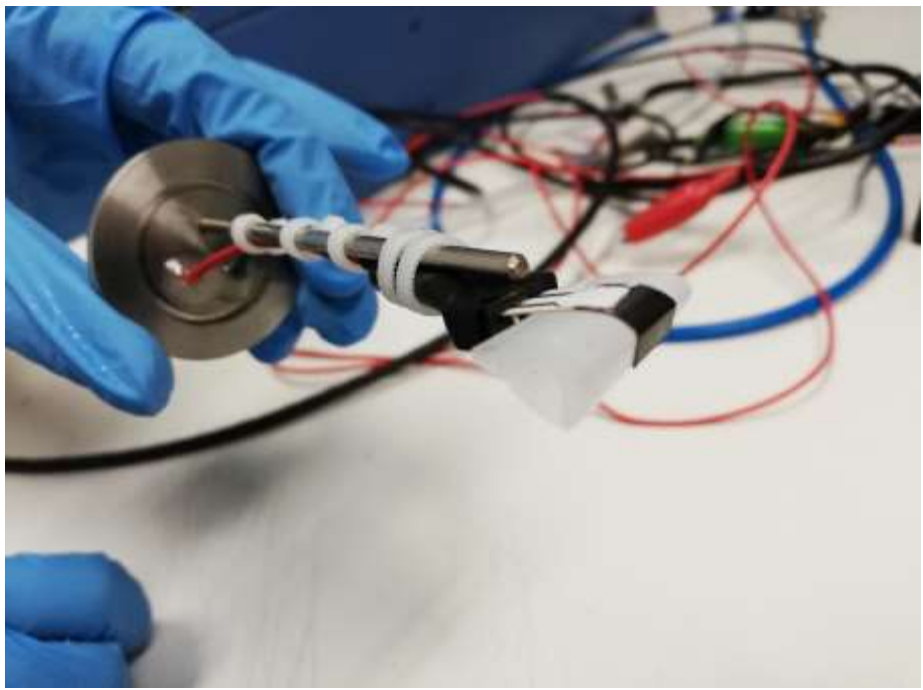

**(a)**

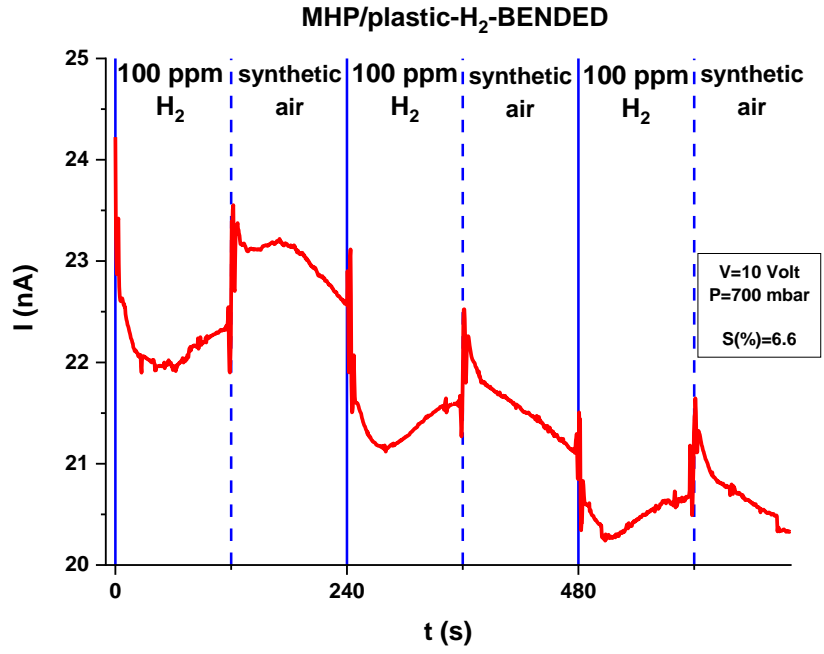

(b)

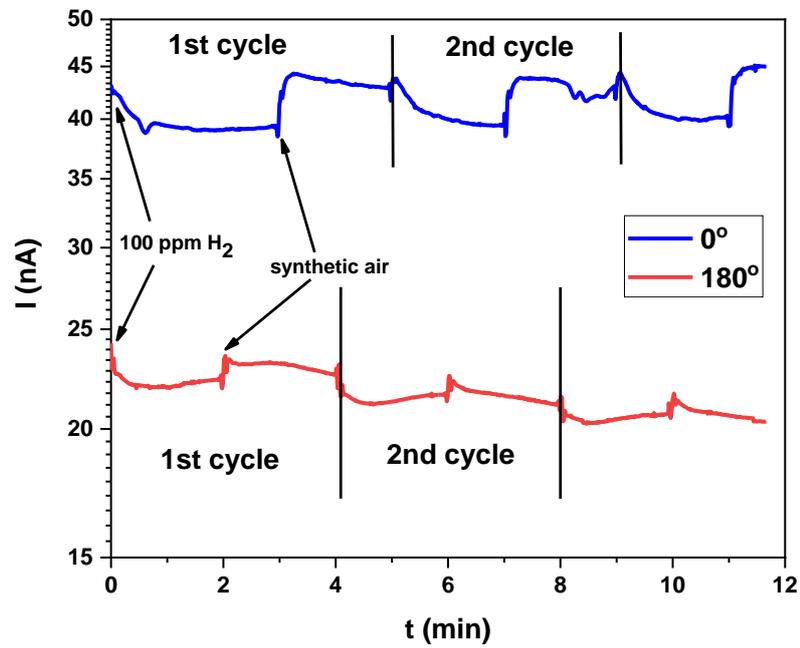

(c)

**Figure 12: (a)** Demonstration of hydrogen sensing ability onto PET based substrate under 10 Volts biasing voltage for 100 ppm H₂ molecule concentration under bending angle of 60°; **(b)**, the sensing element bended up to 180° **(c)**

## 3. Conclusions

We report here for the first time, solution processed hybrid mixed halide perovskite films ($CH_3NH_3PbI_{3-x}Cl_x$) as gas sensor elements. The films were prepared and spin coated onto rigid (glass) and flexible (PET), prepatterned with interdigitated electrodes, substrates. All films operated as self-powered sensing elements for $H_2$ at room temperature. Their sensing properties were based on modulations of its electrical resistance. $H_2$ sensing measurements revealed very promising results regarding the potential of this material as $H_2$ sensing element; the p-type semiconductor characterised by maximum sensitivities of the order of 5-7%, with very fast reaction times of the order of few seconds and the capability to be able to distinguish between two different tested gases ($H_2$ and $O_3$). The compatibility of the demonstrated sensing element with flexible substrates was also confirmed. Ageing measurements showed that the devices retained their sensing abilities even after three weeks of storage. However, despite the first promising results obtained, further work is planned be done in order to improve mixed halide perovskite film's sensing performance towards $H_2$ gas molecules. First of all, the operational mechanism should be confirmed and studied further. Second, doped hybrid perovskite mixed halide films or nanocomposites with graphene-based materials that exhibit higher conductivities, should be tested. The impact of higher conductivities to the sensing performance must be linked. This is expected to enhance the acquired sensitivity. Another idea, is to try to apply 2D solution processed perovskite materials since the lowering of dimension leads to higher surface to volume ratio and thus to the enhancement of all the figures of merit of a sensor: (a) higher sensitivities; (b) broader limit of detection; (c) higher stabilities. The better engineering of mixed halide perovskite probably will provide more tolerant to ion migration problem and will allow the application of higher applied voltages. This is expected to elevate the sensitivity of the particular sensing elements. Stimulated emission Raman Spectroscopy principles will be employed as a tool to be able to distinguish the various gases in situ.

Based in the above, it may be concluded that the hybrid mixed halide perovskite films could be a promising self-powered p-type sensing material for reducing gases as $H_2$ at room temperature.

**Acknowledgements**

Part of this work was supported by the project "National Research Infrastructure on nanotechnology, advanced materials and micro/nanoelectronics" (MIS 5002772), as well as by the action "QUALITY of LIFE" (MIS 5002464) both of which are implemented under the "Action for the Strategic Development on the Research and Technological Sector" funded by the Operational Programme "Competitiveness, Entrepreneurship and Innovation" (NSRF 2014-2020) and co-financed by Greece and the European Union (European Regional Development Fund).

**Corresponding Author**

*Address correspondence to c.petridischania@gmail.com

**Notes**

The authors declare no competing financial interest.